\documentstyle[aps,multicol,epsf]{revtex}

\begin{document}
\draft
\title{Nanowire formation on sputter eroded surfaces}
\author{J. Kim,$^1$ B. Kahng,$^1$ and A.-L. Barab\'asi$^2$}
\address{$^1$ School of Physics and Center for Theoretical Physics,\\ 
Seoul National University, Seoul 151-747, Korea\\ 
$^2$ Department of Physics, University of Notre Dame, 
Notre Dame, IN 46556\\} 
\maketitle
\date{today}

\begin{abstract}
Rotated ripple structures (RRS) on  sputter eroded surfaces
are potential candidates for nanoscale wire fabrication.
We show that the necessary condition for RRS formation is
that the width of the collision cascade in the longitudinal direction
has to be larger than that in the transverse direction,
which can be achieved by using high energy ion beams.
By calculating the structure factor for the RRS
we find that they are more regular and their amplitude is more enhanced
compared to the much studied ripple structure forming
in the linear regime of sputter erosion.
\end{abstract}

\begin{multicols}{2}
The fabrication of nanoscale surface structures
such as quantum dots (QDs) and quantum wires (QWs),
have attracted considerable attention
due to their applications in optical and electronic devices \cite{review}.
These nanostructures form thanks to various self-assembled mechanisms,
induced by the combined effect of strain and growth kinetics.
Yet, the strained nanostructures obtained by these methods
have a size distribution wider than required by applications,
and display random alignment.
Lithographic methods \cite{reed} are often considered
prime candidates to overcome these shortcomings,
but their limited resolution offers further challenges.
Consequently, there is continued high demand for alternative methods 
that would allow low cost and efficient mass fabrication of nanoscale 
surface structures.
In the light of these technological and scientific driving forces,
the recent demonstration by Facsko {\it et al}. that low-energy 
(40 eV $\sim$ 1.8 keV) normal incident Ar$^+$ sputtering 
on GaSb (100) surfaces leads to nanoscale islands
which display remarkably good hexagonal ordering
and have a uniform size distribution,
has captured the interest of the scientific community \cite{sputtering,hexa}.
 
It is known that the morphological evolution of a sputter eroded surface 
is well approximated by the noisy nonlinear Kuramoto-Sivashinsky (KS) equation, 
\begin{eqnarray}
{\partial_t h} &=& \nu_x \partial_x^2 h +\nu_y \partial_y^2 h
-D_{xx}\partial_x^4 h -D_{yy}\partial_y^4 h-D_{xy}\partial_x^2
\partial_y^2 h \nonumber \\
& &+{\lambda_x \over 2}(\partial_x h)^2 + {\lambda_y \over 2}
(\partial_y h)^2+\xi(x,y,t),
\label{ks}
\end{eqnarray}
where $\nu_x$ and $\nu_y$ are the effective surface tensions generated 
by the erosion process; $D_{xx}$, $D_{yy}$, and $D_{xy}$ are the ion 
induced effective diffusion constants; $\lambda_x$ and $\lambda_y$ 
describe the tilt-dependent erosion rates in each direction; 
and $\xi(x,y,t)$ is an uncorrelated white noise with zero mean, 
mimicking the randomness resulting from the stochastic nature
of ion arrival to the surface \cite{cuerno,maxim1}.
At low temperatures all the coefficients in Eq.~(\ref{ks}) depend on
experimental parameters such as the ion beam flux $f$,
the ion beam energy $\epsilon$,
and the incidence angle of ion beam $\theta$ \cite{maxim2}, 
while at high temperatures $D$ depends on surface temperature.
For the sputter erosion process $\nu <0$ and $D >0$,
while the signs of $\lambda_x$ and $\lambda_y$ vary
depending on the incident angle of the ion beam. 

Recently numerical simulations have shown that
there is a clear separation of the linear and nonlinear regimes in time
\cite{spark}: 
Up to a crossover time $\tau_1$ the surface is eroded 
as if the nonlinear terms would be completely absent, 
following the predictions of the linear theory \cite{harper}
(i.e. $\lambda_x = \lambda_y = 0$ in Eq.~(\ref{ks})). 
After $\tau_1$, however, the nonlinear terms
with coefficients $\lambda_x$ and $\lambda_y$
take over and completely determine the surface morphology.
The transition from the linear to the nonlinear regime
can be seen either by monitoring the surface width
(which is proportional to the ripple amplitude)
or the erosion velocity. 

In the nonlinear regime
the case $\lambda_x \lambda_y <0$ is in particular interesting.
The surface morphology in this case exhibits another transition
from kinetic roughening to a rotated ripple structures (RRS)
at a second crossover time $\tau_2$ ($\tau_2 > \tau_1$),
as first predicted by Rost and Krug \cite{krug},
and observed numerically in \cite{spark}.
Moreover, it was found that
the RRS is straighter than the ripple pattern forming in the linear regime, 
making it a potential candidate for nanowire fabrication. 
In this Letter we investigate the necessary conditions
for the formation of RRSs and the impact of the various experimentally
controllable parameters on the morphology of the RRS.
We find that at low temperatures the RRSs form when
the longitudinal width ($\sigma$) of the damage cascade generated
by the ion beam is larger than the transverse width ($\mu$), which 
can be achieved for high energy ($\epsilon$).
When the energy is too high, however,  
the crossover time $\tau_2$ becomes too long.
Thus, we predict that the RRS structure could be 
obtained in a moderate range of the incident energy,
using an appropriate exposure time and low temperatures. 


Rotated ripple structures form when $\lambda_x \lambda_y <0$ and
the rotation angle is given by $\phi_c =\tan^{-1}\sqrt{-\lambda_x/\lambda_y}$.   
As shown in Fig.~1, the angle $\phi_c$ increases with the ratio $a_{\mu}=a/\mu$,
but decreases with the incident angle $\theta$,
where $a$ denotes the penetration depth of the ion beam and 
$a_{\mu}$ depends on the ion beam energy $\epsilon$. 
The KS equation in the rotated frame can be written in the same form
as Eq.~(\ref{ks}) except that the coefficients
$\nu$, $D$, and $\lambda$ are replaced by $\nu'$, $D'$, and $\lambda'$, 
which are functions of those in the original frame and the angle $\phi_c$.
In the rotated frame one of the coefficients of the nonlinear terms,
say $\lambda'_{x'}$, is equal to zero,
and the other is given by $\lambda'_{y'}=\lambda_x+\lambda_y$, 
where $(x',y')$ represents the coordinates in the rotated frame.
Since $\lambda'_{x'}$ vanishes, the dynamic equation 
in the $x'$ direction becomes linear.
Consequently, the ripple pattern is along the $x'$ direction
as long as (i) $\nu_{x'}' <0$; (ii) $D_{x'x'}' >0$; and  
(iii) $\lambda_x \lambda_y <0$. Therefore the conditions (i), (ii), 
and (iii) are the necessary conditions for the formation of the RRS. 

We investigate the satisfiability of these conditions in the parameter space 
($\theta,a_{\mu}= a/\mu$) for different values of $a_\sigma=a/\sigma$. 
We find that when $a_{\mu}>1$ and $a_{\sigma}=1$
or $a_{\mu}>2$ and $a_{\sigma}=2$,
the RRS can form in the region depicted in Fig.~2. 
For $a_{\mu}<1$ given $a_{\sigma}=1$,
the shaded region satisfying  (i) -- (iii) scarcely exists,
so that the formation of RRS is less likely.
That means the RRSs are expected to form when the longitudinal width $\sigma$
is larger than the transverse width $\mu$, 
that is, $\sigma > \mu$. 
Recent experimental results indicate that 
for graphite surfaces $\sigma$ depends on $\epsilon$, 
while $\mu$ is independent of $\epsilon$ for large $\epsilon$ 
($2 \sim 50$ keV) \cite{damage}.
Therefore, the $\sigma > \mu$ condition can be met
when the energy of the incident ion beam is high enough.
Thus in order to obtain the RRS experimentally
it is desirable to use high energy ion beam
with an appropriate choice of the incident angle (Fig.~2). 
However, the use of a high energy ion beam increases $\tau_2$ rapidly,
requiring a longer exposure time.  

Since the nonlinear term disappears in the $x'$ direction
the surface in this direction is driven by a linear instability.
The amplitude of the RRS grows exponentially with time
until the nonlinear term in the $y'$ direction becomes effective,
after which the amplitude saturates. 
Meanwhile, the surface in the $y'$ direction displays 
kinetic roughening due to the presence of the nonlinear 
term $\lambda'_{y'} (\partial_{y'} h)^2 / 2$, 
so that the roughness in the $y'$ direction is considerably 
reduced compared with the roughness in the $x'$ direction.
Therefore the RRS develops a rough morphology in the $x'$ direction,
while it is relatively smooth in the $y'$ direction,
the end configuration resembling a V-shaped wire pattern,
as shown in Fig.~3 \cite{vgroove}. 
The RRS formed in the nonlinear regime is comparable with 
the ripple pattern formed in the linear regime, where there 
are modulations in both directions, and 
the roughness in each direction is almost of the same order. 

We also examined the structure factor
\begin{equation}
S({\bf q})=\int {d{\bf r}\over {(2\pi)^2}}\exp(i {{\bf q}\cdot {\bf r}}
) H({\bf r}),
\end{equation}
where $H({\bf r})$ is the height-height correlation function, 
\begin{equation}
H({\bf r},t)= \left<\sum_{\bf r'}h({\bf r'}+{\bf r},t)h({\bf r'},t) \right>
- \left< \sum_{\bf r'}h^2({\bf r'},t) \right> ,
\end{equation}  
averaged over different configurations \cite{sinha}. 
We find that the structure factor exhibits a peak at $(q_{x,c},q_{y,c})$,
corresponding to ($q'_{x',c}, 0$) in the rotated coordinaates (Fig.~4).
The fact that $q'_{y',c}=0$ 
confirms that the RRS is straight in the $y'$ direction. 
Moreover, the amplitude of the structure factor for the RRS is 
much larger compared to the ripple formed in the linear regime,
as shown in the inset of Fig.~4.
Accordingly, the RRS could be a good candidate for
the fabrication of nanowires.
 
Applying the linear instability theory in the $x'$ direction
we obtain the wavelength of the RRSs
$\ell'_{x'}=2\pi \sqrt{{2 {D'_{x'x'}}}/{|\nu'_{x'}|}}$.
The wavelength depends on the penetration depth linearly,
i.e. $\ell'_{x'} \sim a$, independent of the ion flux $f$. 
In general, the penetration depth depends 
on the incident ion energy $\epsilon$ as $a\sim \epsilon^{2m}$, 
where $m\approx 1/4$ was obtained recently for the nanoscale 
dot structure in low energy sputtering \cite{energy}. 
We also examined $\ell'_{x'}$ in function of 
the ratio $a_{\mu}$ and incident angle $\theta$, observing 
a monotonically increasing and decreasing behavior, respectively (Fig.~5).  

In the high temperature limit the wavelength of the RRS depends
on the ion energy as $\ell'_{x'} \sim \epsilon^{-1/2}$,
and on the ion flux as $\sim f^{-1/2}$,
and on the temperature as $\sim \exp(-\epsilon/4kT)$,
a dependence similar to that observed during the formation 
of nanoscale dots \cite{qdots}. 

In summary, we examined the necessary conditions for the 
formation of RRSs, potential candidates 
for nanowires for electron transport.
We predict that the RRS can be generated under high energy ion beam,
in contrast with the formation of nanoscale dots structure
occurring during low energy ion beam sputtering.
Since high energy ion beam causes longer crossover time
for the formation of the RRS,
exposure time of the ion beam has to be adjusted to obtain the desired RRS.  

This work was supported by ONR,
the Korean Research Foundation (Grant No. 99-041-D00150),
NSF-DMR 01-08494 and NSF-INT 99-10426.

\begin{figure}[ht]
\centerline{\epsfxsize=8.3cm \epsfbox{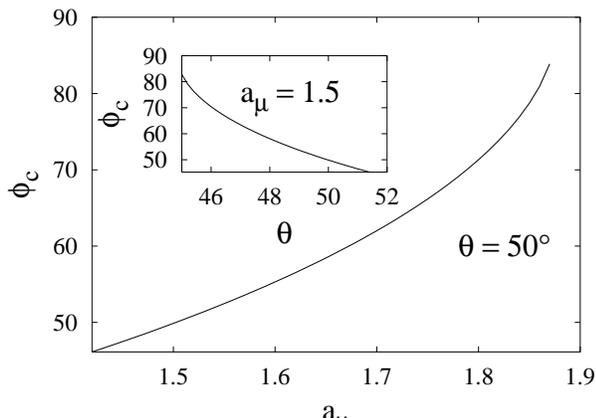}}
\caption{
The rotation angle $\phi_c$ as a function of $a_{\mu}$
at $\theta = 50$ degree.
The inset shows $\phi_c$ versus $\theta$ for $a_{\mu} = 1.5$.
The rotation angle increases as the ratio $a_{\mu}$ increases,
but decreases with the incident angle $\theta$.
}
\end{figure}

\begin{figure}[ht]
\centerline{\epsfxsize=8.3cm \epsfbox{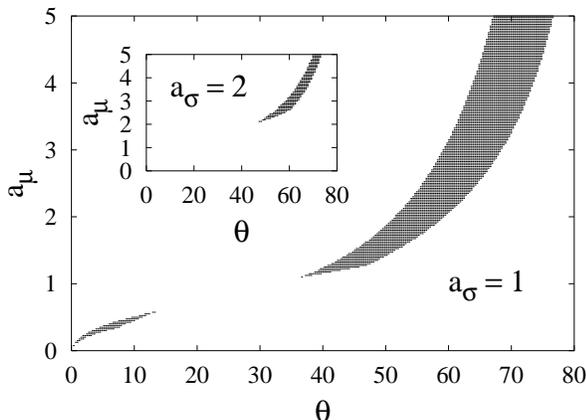}}
\caption{The shaded region in the parameter space $(\theta,a_{\mu})$ 
for $a_{\sigma} = 1$ (inset: for $a_{\sigma} = 2$) corresponds to
the region where the RRS can form.}
\end{figure}

\begin{figure}[ht]
\centerline{\epsfxsize=6.0cm \epsfbox{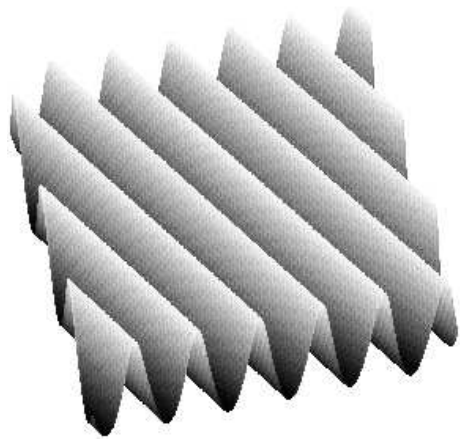}}
\caption{
Surface morphology of the RRS, as generated by numerical simulations,
with $a_{\mu} = 1.3$, $a_{\sigma} = 1$ and $\theta = 43.56$ degree.
}
\end{figure}

\begin{figure}[ht]
\centerline{\epsfxsize=8.3cm \epsfbox{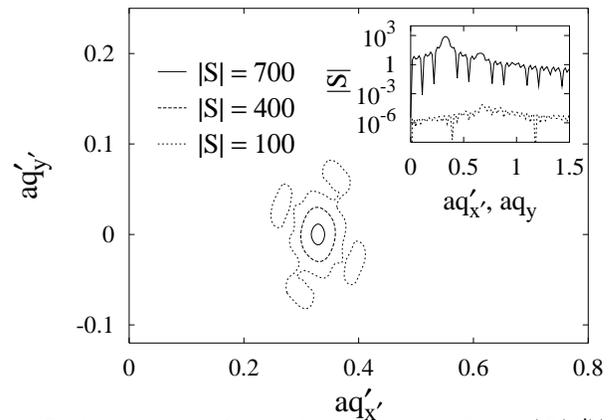}}
\caption{
The amplitude of the structure factor $|S({\bf q'})|$ for the RRS 
shown in Fig.~3. The peak of the structure factor is at 
$aq'_{x',c}=0.33$ and $q'_{y',c}=0$,
implying that the wire structure is straight along the $y'$ axis.
The inset shows the comparison between the amplitudes of the 
structure factor for the RRS (solid line) and for the linear 
ripple (dotted line), implying the amplitude of the RRS is about 
a factor of $10^7$ larger compared to that of the ripple 
structure formed in the linear regime.
}
\end{figure}

\begin{figure}[ht]
\centerline{\epsfxsize=8.3cm \epsfbox{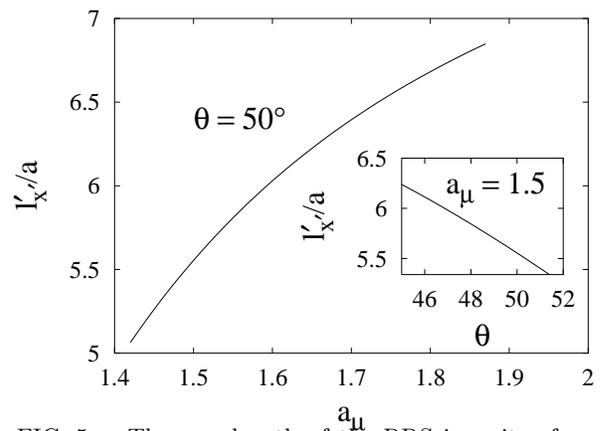}}
\caption{
The wavelength of the RRS in units of $a$ as a,
showing function of $a_{\mu}$ ($\theta = 50$ degree).
Inset: The wavelength versus $\theta$ at $a_{\mu} = 1.5$.
}
\end{figure}

\end{multicols}
\end{document}